\documentclass[twocolumn,english,aps,prx,superscriptaddress]{revtex4-1}
\usepackage[T1]{fontenc}
\usepackage{color}
\usepackage{amsmath}
\usepackage{amssymb}
\usepackage{graphicx}
\usepackage{natbib}
\usepackage{braket}
\usepackage{psfrag}
\usepackage{tikz}
\usepackage{physics}
\usepackage{cancel}
\usepackage{bbm}

\usepackage[colorlinks=true,urlcolor=blue,citecolor=blue,linkcolor=blue]{hyperref}

\usepackage{babel}

\usepackage[normalem]{ulem}

\DeclareMathOperator*{\argmin}{argmin}

\usepackage{xcolor}

\usepackage{array,booktabs}
\newcolumntype{M}[1]{>{\centering\arraybackslash}m{#1}}

\begin{document}

\title{Characterization and tomography of a hidden qubit}

\author{M. Pechal}
\altaffiliation{Current address: Department of Physics, ETH Zurich, CH-8093 Zurich, Switzerland}
\affiliation{IBM Quantum, IBM Research -- Zurich, S\"aumerstrasse 4, 8803 R\"uschlikon}
\author{G. Salis}
\affiliation{IBM Quantum, IBM Research -- Zurich, S\"aumerstrasse 4, 8803 R\"uschlikon}
\author{M. Ganzhorn}
\affiliation{IBM Quantum, IBM Research -- Zurich, S\"aumerstrasse 4, 8803 R\"uschlikon}
\author{D. J. Egger}
\affiliation{IBM Quantum, IBM Research -- Zurich, S\"aumerstrasse 4, 8803 R\"uschlikon}
\author{M. Werninghaus}
\affiliation{IBM Quantum, IBM Research -- Zurich, S\"aumerstrasse 4, 8803 R\"uschlikon}
\author{S. Filipp}
\affiliation{IBM Quantum, IBM Research -- Zurich, S\"aumerstrasse 4, 8803 R\"uschlikon}
\affiliation{Technical University Munich, Department of Physics, 85748 Garching, Germany}

\begin{abstract}
In circuit-based quantum computing the available gate set typically consists of single-qubit gates acting on each individual qubit and at least one entangling gate between pairs of qubits. In certain physical architectures, however, some qubits may be 'hidden' and lacking direct addressability through dedicated control and readout lines, for instance because of limited on-chip routing capabilities, or because the number of control lines becomes a limiting factor for many-qubit systems. In this case, no single-qubit operations can be applied to the hidden qubits and their state cannot be measured directly. Instead, they may be controlled and read out only via single-qubit operations on  connected 'control' qubits and a suitable set of two-qubit gates. We first discuss the impact of such restricted control capabilities on the quantum volume of specific qubit coupling networks. We then experimentally demonstrate full control and measurement capabilities in a superconducting two-qubit device with local single-qubit control and iSWAP and controlled-phase two-qubit interactions enabled by a tunable coupler. We further introduce an iterative tune-up process required to completely characterize the gate set used for quantum process tomography and evaluate the resulting gate fidelities.
\end{abstract}

\date{\today}

\maketitle

\section{Introduction}
The sizes of engineered quantum systems encountered in state-of-the-art laboratories \cite{Krantz2019,Castelvecchi2017,Arute2020,Wright2019,Jurcevic2020} have been steadily increasing, enabled by progress in packaging technology \cite{Rosenberg2017} and integration of control electronics \cite{ZIQuantumComputingControlSystem}. Scaling such systems even further, however, still raises practical challenges as the amount of required control hardware and signal lines is proportional to the growing number of qubits. A less well explored approach to scaling, complementary to control electronics integration, is to reduce the number of control lines per qubit.

Traditionally, superconducting circuit systems are designed with individual charge and flux controls for each qubit \cite{Andersen2020} or with charge controls for qubits and flux controls for qubit couplers \cite{McKay2016}. Here, we focus on the latter control approach which for large linear chains or square lattice arrangements with nearest neighbor couplings results in approximately two or three input lines per qubit, respectively.
A more favorable ratio of control lines per qubit may be achieved by forgoing the direct control lines of a fraction of the qubits and controlling them indirectly by means of their coupling to neighboring qubits. We call such qubits \emph{hidden} and as we show here, they allow one to reduce the total number of control lines without compromising the computational power of the device.

In setups using multiplexed measurement schemes \cite{Chen2012f,Heinsoo2018}, readout only accounts for a small fraction of the total microwave line count, so the potential for improvement by reducing the number of readout lines may at first sight seem more limited than for control lines. In these layouts, a single feed line is shared between multiple readout resonators. Nevertheless, each qubit needs to be coupled to its own readout resonator, taking up a significant fraction of the area on the chip. Reducing the number of qubits with direct readout may therefore still be beneficial as it would allow a more economical use of the chip area. We therefore consider hidden qubits that lack not only direct control but also direct readout.

The lack of direct addressability of some subsystems is also an inherent feature in certain devices such as quantum memories using multi-mode microwave circuits \cite{Naik2017} or nanomechanical resonators \cite{Pechal2018}. Also silicon spin qubits \cite{Zajac2018,Watson2018}
would strongly benefit from the development of indirect control and readout techniques \cite{Sigillito2019}
because of the envisaged dense integration \cite{Li2018}
where exchange-type coupling between qubits can be controlled directly by a gate voltage, but read-out and single-qubit control require additional complexity.

Here, we first discuss quantum computing architectures making use of hidden qubits and analyze a specific two-dimensional grid configuration as an example. We quantify its computational power by estimating the quantum volume \cite{Moll2018,Cross2019} and show that for a given number of controls, a higher quantum volume can be reached when hidden qubits are used, assuming that error rates are reduced at least one order of magnitude below current state-of-the-art.

We further present an experimental demonstration of full control and readout of a superconducting qubit system with one control and one hidden qubit. We show that despite the lack of direct control and readout of one of the qubits, we can reliably tune up all the gates necessary to fully control and measure the two-qubit system.

We also develop a modified quantum process tomography method which allows us to characterize the gates in a way that is robust against state preparation and measurement errors. This is especially important for systems with hidden qubits where preparation and measurement of arbitrary states relies heavily on the use of typically more error-prone two-qubit gates but our technique can in principle also be applied to standard network topologies without hidden qubits.

\section{Configurations with hidden qubits}

\begin{figure}
\includegraphics[width=8.5cm]{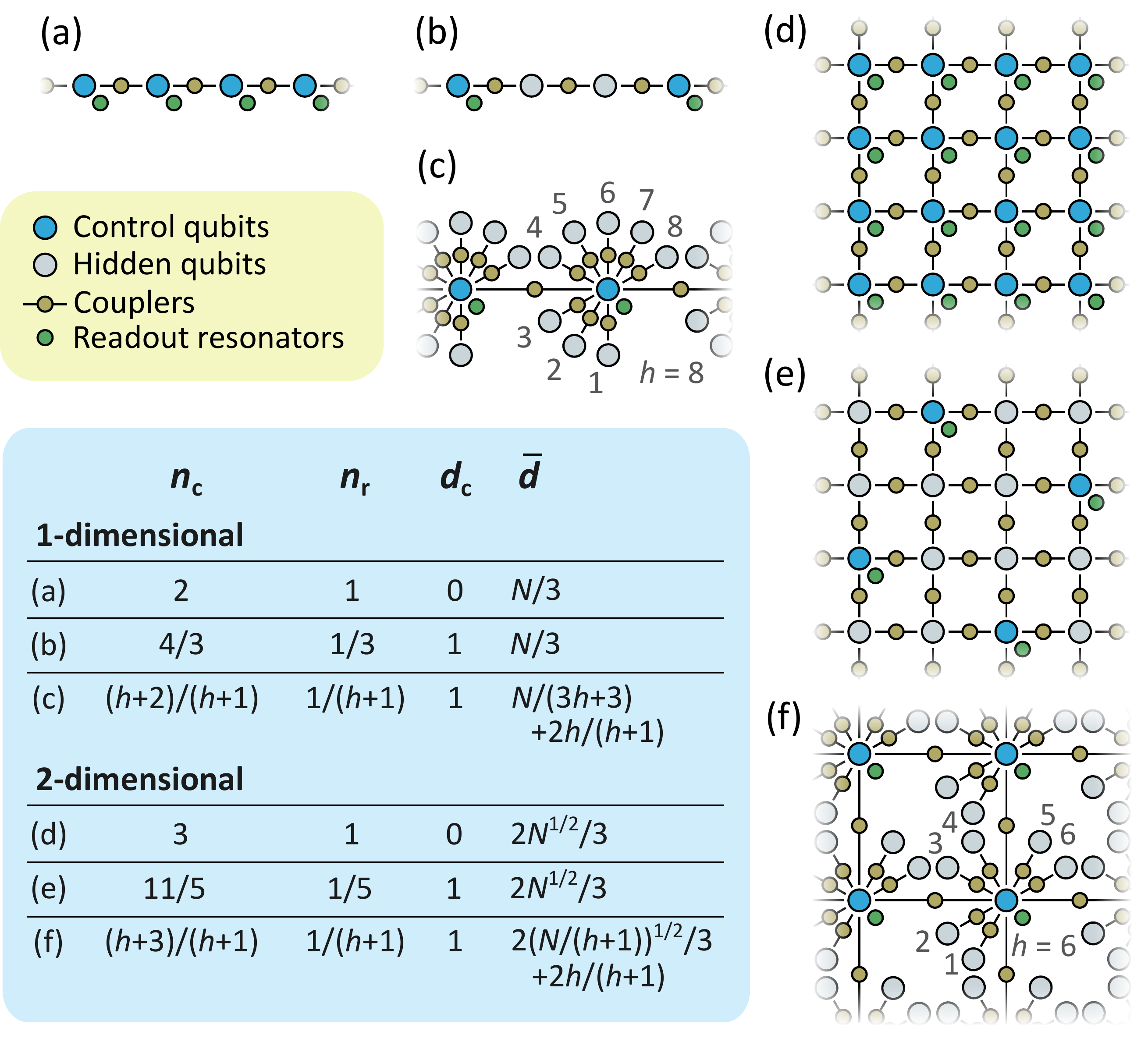}
\caption{Illustration of a few different 1-dimensional (a,b,c) and 2-dimensional (d,e,f) qubit network configurations. (a) and (d) represent the standard setting with direct control and readout of each qubit. (b) and (e) are similar configurations but with a fraction of the qubits hidden. This fraction is chosen such that each hidden qubit is adjacent to a control qubit. (c) and (f) are settings where each control qubit has a number $h$ of hidden qubits attached to it in a star geometry. The number of control lines per qubit $n_\mathrm{c}$, the number of readout resonators per qubit $n_\mathrm{r}$, the maximal distance to a control qubit $d_\mathrm{c}$ and the average distance $\overline{d}$ between two randomly chosen qubits as a function of the total number of qubits $N$ are listed for comparison of the different configurations.}
\label{fig:networks}
\end{figure}

To illustrate the reduction in readout and control lines we consider different qubit networks as shown in Fig.~\ref{fig:networks}. The linear chain and the square lattice with direct control and readout of all qubits shown in Fig.~\ref{fig:networks}(a,d) serve as our baseline to evaluate configurations with hidden qubits. The numbers of control lines $n_{\mathrm{c}}$ per qubit (in the limit of a large number of qubits $N$) for the linear chain and the square lattice are $n_{\mathrm{c}}=2$ (one control for each qubit and one for each coupler) and $n_{\mathrm{c}}=3$ (one control for each qubit and two for each coupler), respectively. In both cases, there is $n_{\mathrm{r}} = 1$ readout resonator per qubit. Each qubit is directly controllable and measurable, which we express using the maximum distance $d_{\mathrm{c}}$ from any qubit to the nearest control and read-out qubit, in these cases trivially $d_{\mathrm{c}} = 0$. Another important parameter that characterizes the network's connectivity is the average distance $\overline{d}$ between a pair of randomly chosen qubits. For the linear chain and the square lattice, this distance can be approximated to leading order in $N$ as $\overline{d} = N/3$ and $\overline{d} = 2N^{1/2}/3$, respectively.

In our discussion of networks with hidden qubits, we restrict ourselves to configurations in which each hidden qubit is at most a distance 1 away from a control qubit. That is, $d_{\mathrm{c}} = 1$. We consider two types of such networks. In the first network, shown in Fig.~\ref{fig:networks}(b,e), we convert a certain fraction of the controllable qubits from Fig.~\ref{fig:networks}(a,d) into hidden qubits. If we wish to satisfy the condition $d_{\mathrm{c}} = 1$, we can have at most two hidden qubits per control qubit in the one-dimensional chain and four hidden qubits per control qubit in the two-dimensional grid. The number of controls (single-qubit drives as well as coupler drives) per qubit is then reduced to $n_{\mathrm{c}}=4/3$ and $n_{\mathrm{c}}=11/5$, respectively. This reduction of roughly 30\% is not exceptionally large because the number of couplers per qubit is unchanged in this configuration. The number of readout resonators per qubit is however lowered quite substantially to $n_{\mathrm{r}}=1/3$ and $n_{\mathrm{r}}=1/5$. The connectivity of the network and therefore also the average distances $\overline{d}$ between random pairs of qubits are the same as for the fully controlled networks from Fig.~\ref{fig:networks}(a,d).

In the second type of network we consider, shown in Fig.~\ref{fig:networks}(c,f), the control qubits have the same connectivity as in the fully controlled configurations from Fig.~\ref{fig:networks}(a,d) but each of them has a number $h$ of hidden qubits coupled directly to it. In such networks, the control qubits form bottlenecks because they have to mediate all interactions between hidden qubits. Moreover, the need to couple a large number of hidden qubits to a single control qubit without introducing unwanted direct couplings may pose additional RF engineering challenges. Nevertheless, these network configurations compare favorably to the ones from Fig.~\ref{fig:networks}(b,e) in terms of the number of control lines $n_\mathrm{c}$ per qubit (by a factor approaching $3/4$ as $h\to\infty$ in the 1d case and $5/11$ in the 2d case) and readout resonators $n_\mathrm{r}$ per qubit (by a factor of $3/(h+1)$ in the 1d case and $5/(h+1)$ in the 2d case). We therefore choose to focus on them in the subsequent discussion.

To quantify the potential advantanges of a system with hidden qubits, we analyze the two-dimensional qubit networks from Fig.~\ref{fig:networks}(f) in more detail. By varying the parameter $h$ -- the number of hidden qubits per control qubit -- we can compare the standard two-dimensional grid ($h=0$) with systems in which hidden qubits dominate ($h\gg 1$). We estimate the achievable quantum volume $V_Q$ \cite{Cross2019} as a function of the number of controls for various values of $h$, assuming that the scaling of control lines and associated resources constrains the size of practical devices. To calculate $V_Q$, we estimate the number of elementary two-qubit gates (taking into account the limited connectivity of the network which also depends on $h$) and the amount of time
necessary to implement average quantum circuits of a given depth. The algorithm we use to do this is described in Appendix \ref{app:VQ}. To get an estimate of $V_Q$, we make the following simple assumptions: the errors in the system are dominated by decoherence (while control errors are neglected) and the overall error probability can be estimated as $\Gamma N T$, where $\Gamma$ is an effective error rate per qubit, $N$ is the number of qubits, and the total duration of the circuit is $T$.  We then analyze the results for $V_Q$ as a function of a natural dimensionless parameter describing errors in the circuit: the error probability per qubit $\Gamma \tau$ in the amount of time $\tau$ taken by a typical (two-qubit) gate.

Current state-of-the-art superconducting qubit systems achieve coherence times around $50\,\mu\mathrm{s}$ while two-qubit gates take on the order of $200\,\mathrm{ns}$ \cite{Kjaergaard2020,Rol2019,Arute2019}. This means that for a realistic decoherence-limited system, we have $\Gamma \tau \approx 0.004$. One could argue that since a two-qubit gate involves the evolution of a pair of qubits, its error due to decoherence should be on the order of $2\Gamma\tau$, in our case $0.008$. This is consistent with current best two-qubit gate fidelities above 99\% \cite{Kjaergaard2020,Negirneac2020,Ganzhorn2020,Foxen2020}.

\begin{figure}
\includegraphics[width=8.0cm]{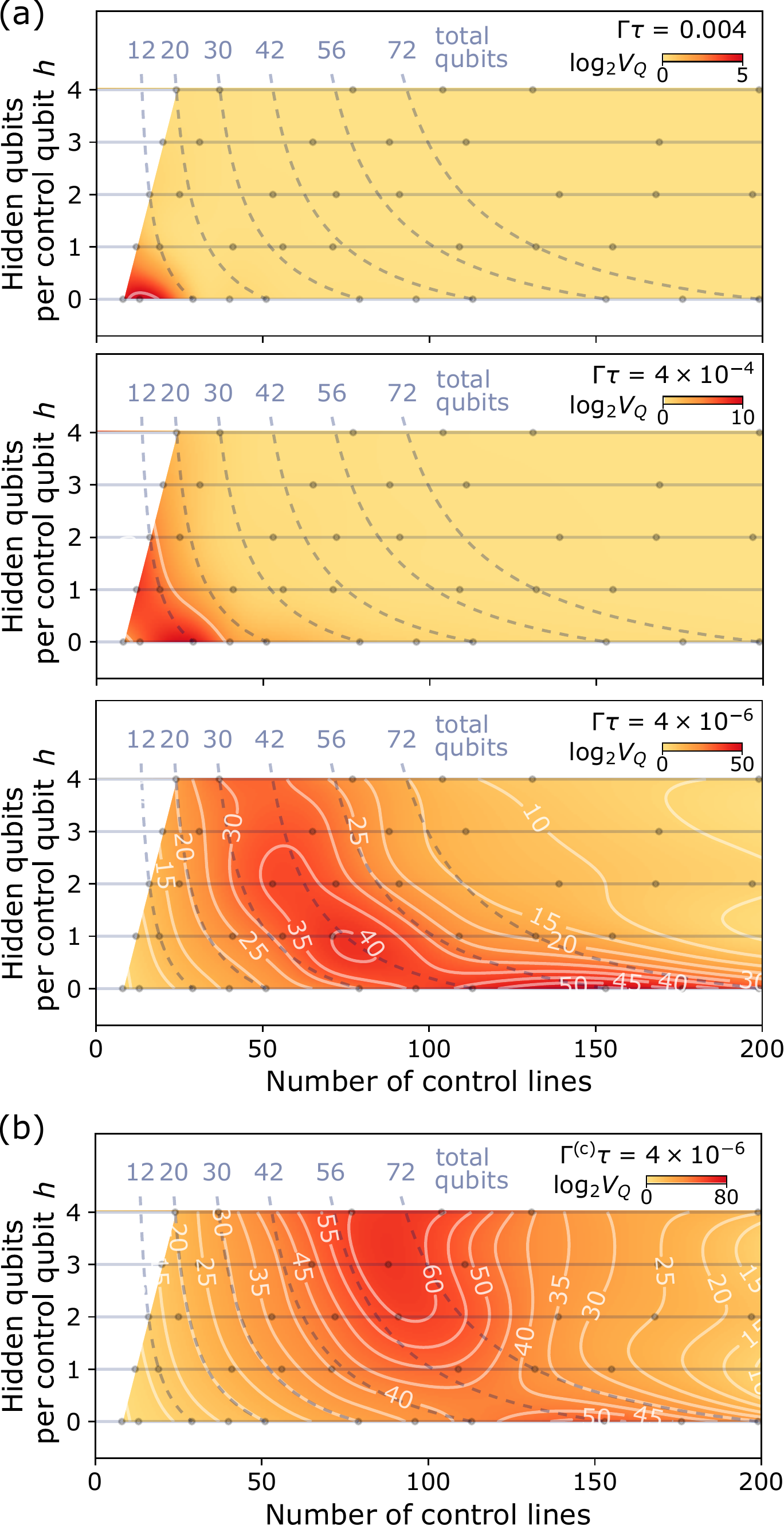}
\caption{Estimated quantum volume as a function of the number of control lines for different numbers $h$ of hidden qubits per control qubit.
The grey dots mark the set of grids for which the quantum volume was calculated. The value indicated by the density plot and the contours was obtained by interpolation from these.
The curved dashed lines indicate constant \emph{total} numbers of qubits.
In (a), we assume all qubits are subject to the same error probability per two-qubit gate duration $\Gamma\tau$ and compare three different values of this parameter ($0.004$, $4\times 10^{-4}$ and $4\times 10^{-6}$). In (b), we assume only control qubits undergo significant decoherence with a probability $\Gamma^{(\mathrm{c})}\tau = 4\times 10^{-6}$ per two-qubit gate duration.}
\label{fig:qvolume}
\end{figure}

The results of the $V_Q$ calculation for $\Gamma\tau=0.004$ and for hypothetical lower values which may be reached with future improvements of quantum hardware are shown in Fig.~\ref{fig:qvolume}.
In the top-most plot, we observe that with currently achievable error probabilities
$\Gamma \tau\approx 0.004$, systems with $h>0$ are not advantageous. That is, at any number of control lines, the quantum volume
is highest on the bottom-most horizontal line, at $h=0$. However, a factor 10 reduction in gate errors, which we believe can be expected in the near future, would lead to configurations with hidden qubits offering a quantum volume on par with standard 2D grid systems (with $h=0$) or better for small numbers of control lines around 20 (middle plot in Fig.~\ref{fig:qvolume}(a)).
Such configurations could therefore be of use in moderately sized quantum devices. Our calculations also indicate that the maximum number of controls at which $h>0$ would be advantageous strongly depends on error rates. Cutting-edge systems with 50-100 controls would start benefiting from a hidden-qubit architecture once error probabilities $\Gamma\tau$ are reduced to the order of $4\times 10^{-6}$ and below. The absence of direct control and readout lines on hidden qubits better isolates them from their environment. Thus, they may reach longer coherence times than control qubits. We therefore also analyze a hypothetical setting in which the decoherence rate of hidden qubits is negligible when compared with control qubits (for which we again take $\Gamma\tau = 4\times 10^{-6}$). The results shown in Fig.~\ref{fig:qvolume}(b) demonstrate that in this setting, systems with $h>0$ provide an advantage for even higher numbers of control lines, up to around 120.
We note that these numbers are only indicative and may vary when considering different network topologies.

\section{Measurement and control}\label{sec:meascontrol}

The gate set required for full control and measurement of a system with hidden qubits is inherently different from the universal gate sets in devices where all qubits are directly accessible. To illustrate this, we focus on a device with one hidden and one control qubit as in our experiment.

To achieve full controllability \cite{Schirmer2001,Romano2006} in two-qubit systems which can implement arbitrary single-qubit operations, only one additional two-qubit gate is needed. Let us assume that this gate is generated by a Hamiltonian $H$ (i.e. its corresponding unitary is $\exp(-\mathrm{i}H t/\hbar)$) and that we can also realize its generalized version with arbitrary other evolution times (rotation angles) $t$. Then the condition for full controllability is that the smallest operator algebra containing $H$ together with $\sigma_i\otimes\mathbbm{1}$ and $\mathbbm{1}\otimes\sigma_i$ (where $i\in\{x,y,z\}$) is the full space $W$ of $4\times 4$ (traceless) Hermitian matrices \cite{Ramakrishna1995}. In other words, the set of arbitrary nested commutators formed from these operators needs to span $W$. This condition is satisfied by a number of two-qubit gates. For instance controlled phase (cPHASE) gates, iSWAP-type and SWAP-type gates, generated by $\sigma_z\otimes\sigma_z$, $\sigma_x\otimes\sigma_x+\sigma_y\otimes\sigma_y$ and $\sigma_x\otimes\sigma_x+\sigma_y\otimes\sigma_y+\sigma_z\otimes\sigma_z$, respectively, all have the required property and therefore form a universal set of gates in combination with single-qubit operations on both qubits.

When we remove the single-qubit rotations of the hidden qubit from the generator set, we find that 
cPHASE- or iSWAP-type gates are no longer sufficient for full controllability. SWAP-type gates, on the other hand, still form a universal gate set and so do the cPHASE- and iSWAP-type gates together. Our experimental setup based on a tunable coupler is better suited for implementations of cPHASE- and iSWAP-type gates rather than a SWAP-type gate and we therefore use them to form the required gate set. Note that in other types of physical systems with full Heisenberg interaction of the form $\sigma_x\otimes\sigma_x + \sigma_y\otimes\sigma_y + \sigma_z\otimes\sigma_z$ (for instance in chains of spin qubits in quantum dot devices \cite{Kandel2019}), the SWAP-type gates would be a more natural choice to implement a universal gate set and realize state transfer within the qubit network \cite{Christandl2004,Christandl2005}.

\begin{figure}
\includegraphics[width=7.5cm]{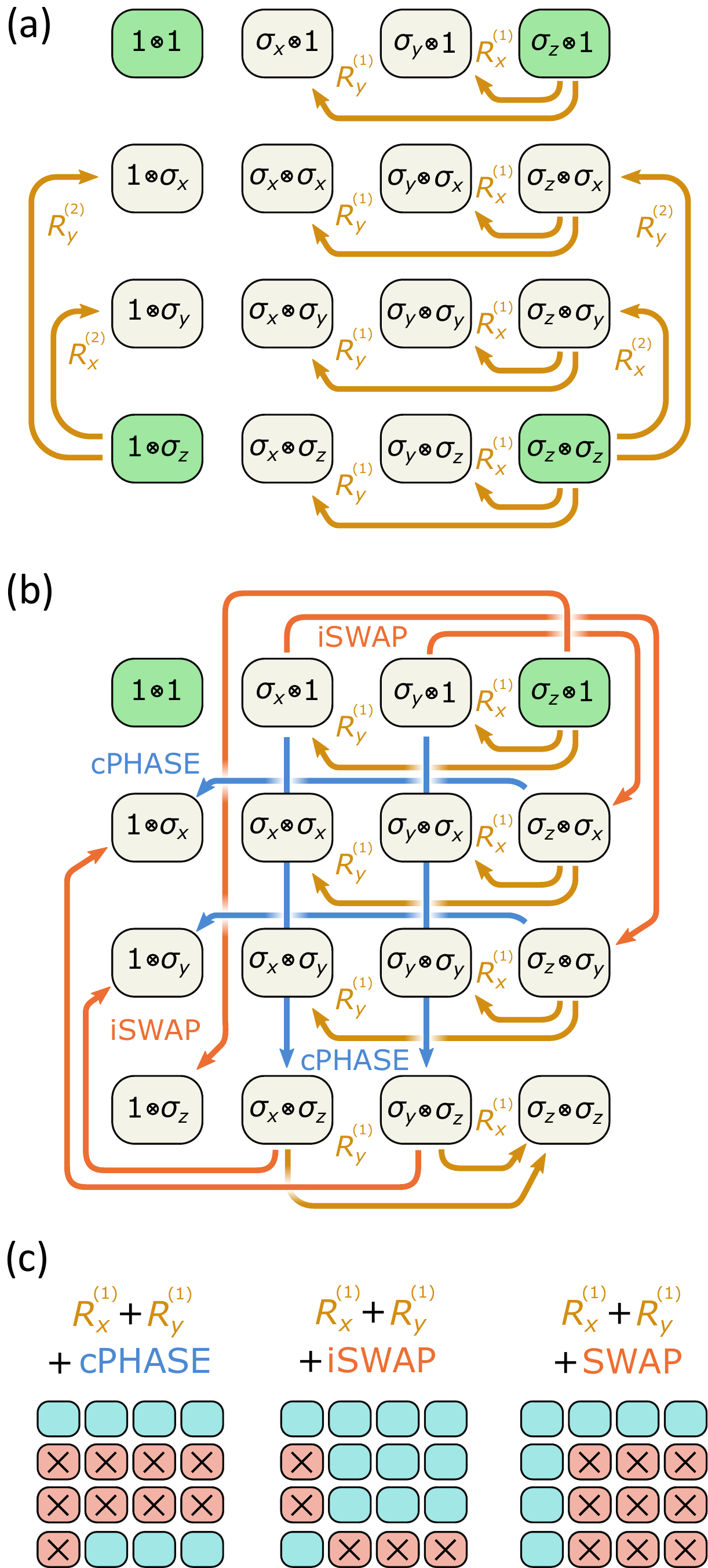}
\caption{Accessibility of the entire space of two-qubit measurement operators (a) in the normal case with full control and readout of both qubits and (b) with the second qubit hidden. The native measurement operators in each case are marked in green while the arrows show transformations of the measurement operators by different gates applied prior to the measurement. $R_x^{(j)}$ and $R_y^{(j)}$ denote $\pi/2$ rotations around $x$ and $y$ applied to qubit $j$. (c) Illustration that single-qubit gates on the control qubit together with one two-qubit gate from the set \{cPHASE, iSWAP, SWAP\} are not sufficient for full two-qubit tomography. The crossed out measurement operators marked in red are inaccessible in each case.}
\label{fig:measOps}
\end{figure}

In addition to full controllability, it is also important to be able to fully measure the state of our system. Joint dispersive readout of both qubits \cite{Filipp2009b} or simultaneous single-shot readout \cite{Walter2017}, commonly used in superconducting systems, is equivalent to measuring the native observables $1\otimes 1$, $1\otimes \sigma_z$, $\sigma_z\otimes 1$ and $\sigma_z\otimes \sigma_z$. Applying unitaries $U_i$ to the final state $\rho_f$ before dispersively measuring the operator $M$ allows us to measure other operators $U_i^{\dagger}M U_i^{\vphantom{\dagger}}$ as the trace is cyclical, i.e. $\mathrm{Tr}[U_i^{\dagger} M U_i^{\vphantom{\dagger}} \rho_f] = \mathrm{Tr}[M U_i^{\vphantom{\dagger}} \rho_f U_i^{\dagger}]$. To perform quantum state tomography and thus fully characterize the state of the system, the set of unitary transformations $U_i$ must be complete in the sense that they map the four native measurement operators onto a set spanning the full space of measurement operators. In the fully controlled case with joint two-qubit readout, this is achievable with only single-qubit rotations, as illustrated in Fig.~\ref{fig:measOps}(a).

If qubit 1 is directly controlled and measured while qubit 2 is hidden, the native measurement operators accessible by a dispersive measurement are only $1\otimes 1$ and $\sigma_z\otimes 1$. It is then no longer sufficient to apply only single-qubit rotations to the state on which we wish to perform tomography. To access the full space of measurement operators, we must extend the set of applied unitaries to contain rotations on qubit 1 as well as two-qubit operations. Fig.~\ref{fig:measOps}(b) shows that the iSWAP and cPHASE gates we have chosen for our universal gate set are also sufficient for full two-qubit tomography. None of these two-qubit gates together with qubit 1 rotations is sufficient by itself (see Fig.~\ref{fig:measOps}(c)). Replacing the iSWAP gate by a SWAP does not make the set complete either, although $\sqrt{\mathrm{SWAP}}$ alone or any pair of two qubit gates from the set \{cPHASE, iSWAP, SWAP\} would suffice.

\section{Experimental system and calibration of hidden qubit operations}

To demonstrate the principle of hidden qubit architectures, we have performed experiments on a fixed-frequency two-qubit system with one control and one hidden qubit. This device is the same as in \cite{Ganzhorn2020}. The control qubit at $6.19~\rm{GHz}$ is coupled to a drive line and a readout resonator while the hidden qubit at $5.09~\rm{GHz}$ interacts only with the control qubit via a parametric tunable coupler. The coupler is operated at a bias point around $7.7~\rm{GHz}$. Both qubits are transmons with anharmonicities of $-290~\rm{MHz}$ and $-310~\rm{GHz}$. Their exchange-type interaction with the coupler has a strength of $116~\rm{MHz}$ and $142~\rm{MHz}$, respectively. The coherence times of the qubits undergo slow variations over time but are most typically around $T_1\approx 30\,\mu\mathrm{s}$, $T_2\approx 30\,\mu\mathrm{s}$ for the control qubit and $T_1\approx 60\,\mu\mathrm{s}$, $T_2\approx 20\,\mu\mathrm{s}$ for the hidden qubit.

The iSWAP gate is realized by parametric driving of the tunable coupler between the two qubits at a frequency close to the difference between the qubits' transition frequencies \cite{McKay2016,Ganzhorn2019,Roth2019,Bengtsson2020}. The drive pulse has a square envelope and its frequency and duration are chosen to obtain maximal excitation transfer from the control to the hidden qubit. This is done by preparing the control qubit in its excited state, applying the iSWAP pulse and measuring the remaining excitation of the control qubit. The length and frequency of the pulse are then adjusted until the final excitation is minimized.

The cPHASE gate is implemented using the transition $|11\rangle \leftrightarrow |20\rangle$ (where the first state refers to the hidden qubit and the second to the control qubit) \cite{Bengtsson2020,Ganzhorn2020}, again induced by parametric driving of the coupler at the appropriate difference frequency. A $2\pi$-rotation in the $\{|11\rangle,|20\rangle\}$ subspace imparts a phase of $\pi$ to the initial $|11\rangle$ state. To set up the drive pulse for this operation, we prepare the system in the $|11\rangle$ state with the previously calibrated iSWAP gate and after applying a square-envelope flux pulse to the coupler, we measure the excitation of the control qubit. Similarly to the setup of the iSWAP gate, we then adjust the length of the pulse to maximize the final population of the control qubit's $|1\rangle$ state.
As in \cite{Ganzhorn2020}, we fine tune the frequency of the $|11\rangle \leftrightarrow |20\rangle$ drive to bring the extra phase accumulated by the $|11\rangle$ state as close to $\pi$ as possible.

In general, the drives enabling the iSWAP and the cPHASE gates also induce single-qubit phases on the individual qubits. To eliminate these phases, each gate is followed by a virtual $Z$ gate \cite{McKay2017}, that is, a shift of the rotating frames associated with the qubits. The appropriate frame shifts are determined by measuring the single-qubit phases in Ramsey-type experiments similar to Ref.~\cite{Ganzhorn2020}, as described in Appendix~\ref{app:gate_set_tuneup}.

Apart from the basic controllability aspects, calibration of the gates needed for full controllability of the system with a hidden qubit is more involved than in the fully controlled setting because of interdependencies between the calibration steps. For instance, if both qubits are directly controllable then characterization of the controlled phase gate requires only single-qubit rotations and joint readout in addition to the gate being characterized. But with a hidden qubit, the superposition states needed to establish the parameters of the cPHASE gate can only be prepared and measured using iSWAP gates. Similarly, some of the parameters of the iSWAP gate such as the phases it imparts to the computational states can only be obtained using pulse sequences containing several iSWAPs (one iSWAP is needed to prepare states with excitation in the hidden qubit and one more for readout). In contrast with the fully controlled case where such parameters can be measured using a single instance of the iSWAP gate together with single-qubit gates and joint readout, this introduces a non-trivial interplay between different parameters of the gate.
Consequently, the parameters of the gate set must be tuned up in a self-consistent way and in the correct order. In our experiment, the single-qubit operations on the control qubit are set up with standard Rabi and Ramsey measurements. The two-qubit gates are then calibrated as described above and in more details in Appendix \ref{app:gate_set_tuneup}.

\section{Process tomography - Demonstration of state preparation and readout}

As shown in Sec.~\ref{sec:meascontrol}, we can perform full state tomography on a system with one control and one hidden qubit with a suitable set of 15 operations consisting of single-qubit gates on the control qubit, an iSWAP and a cPHASE gate. This set is
\begin{align*}
& \mathrm{ID},\\
& \mathrm{R}_j(\pi/2) & \text{for }j\in\{x,y\},\\
& \mathrm{iSWAP},\\
& \mathrm{R}_j(\pi/2).\mathrm{cPHASE} & \text{for }j\in\{x,y\},\\
& \mathrm{R}_j(\pi/2).\mathrm{iSWAP} & \text{for }j\in\{x,y\},\\
& \mathrm{R}_j(\pi/2).\mathrm{iSWAP}.\mathrm{cPHASE} & \text{for }j\in\{x,y\},\\
& \mathrm{R}_k(\pi/2).\mathrm{iSWAP}.\mathrm{R}_j(\pi/2)  & \text{for }j,k\in\{x,y\},\\
& \mathrm{R}_x(\pi/2).\mathrm{cPHASE}.\mathrm{R}_x(\pi/2), &
\end{align*}
which effectively transform the native measurement operator $\sigma_z\otimes\mathbbm{1}$ into all 15 non-trivial two-qubit Pauli operators (see Fig.~\ref{fig:measOps}(b)). Here, we write sequences of gates in the form $G_n .\cdots.G_2.G_1$,
chronologically ordered from right to left. ID is the identity operation (empty gate sequence) and $\mathrm{R}_j(\varphi)$ with $j\in\{x,y\}$ denotes a single-qubit rotation around axis $j$ by an angle $\varphi$.

To prepare a full set of basis states for the two qubits we apply a suitable set of 16 sequences to the initial ground state $|00\rangle$. We use sequences of the form $A_2.\mathrm{iSWAP}.A_1$ (optionally without the iSWAP if $A_1 = \mathrm{ID}$) in any of the 16 possible combinations, where the subsequences $A_{1,2}$ are one of the following four operations  $A_i=\{\mathrm{ID},\, \mathrm{R}_x(\pi),\,\mathrm{R}_x(\pi/2),\,\mathrm{R}_y(\pi/2)\}$ acting on the control qubit. The basis state preparation together with the state tomography procedure allows us to perform quantum process tomography (QPT) \cite{Chuang1997,Poyatos1997} and characterize an arbitrary process $X$. This is based on 240 measurements with sequences $B.X.A$, where $A$ is any of the 16 state preparation sequences and $B$ any of the 15 tomography sequences described above.

The unknown process is reconstructed by solving the least-squares problem
\begin{equation}\label{eq:QPTlstsq}
  \argmin_{\mathcal{X}}
  \sum_{A,B} |
  \mathrm{Tr}\,
  \left[M
  \mathcal{B}\circ\mathcal{X}\circ\mathcal{A}(
  \rho_0)\right]
  -\mu_{A,B}|^2.
\end{equation}
Here, $M = \sigma_z\otimes\mathbbm{1}$ is the native measurement operator, $\rho_0 = |00\rangle\langle 00|$ the initial state and $\mu_{A,B}$ the measurement outcome for the pair of preparation and tomography sequences $A$ and $B$. The symbols $\mathcal{A}$, $\mathcal{B}$ and $\mathcal{X}$ stand for the superoperators describing the sequences $A$ and $B$ and the unknown process $X$. The minimization above is performed under the constraint that $\mathcal{X}$ is a completely positive trace-preserving map. This is equivalent to a semidefinite programming problem \cite{Vandenberghe1996} which we solve using the \texttt{cvxopt} module in Python.

\begin{figure}
\includegraphics[width=8.5cm]{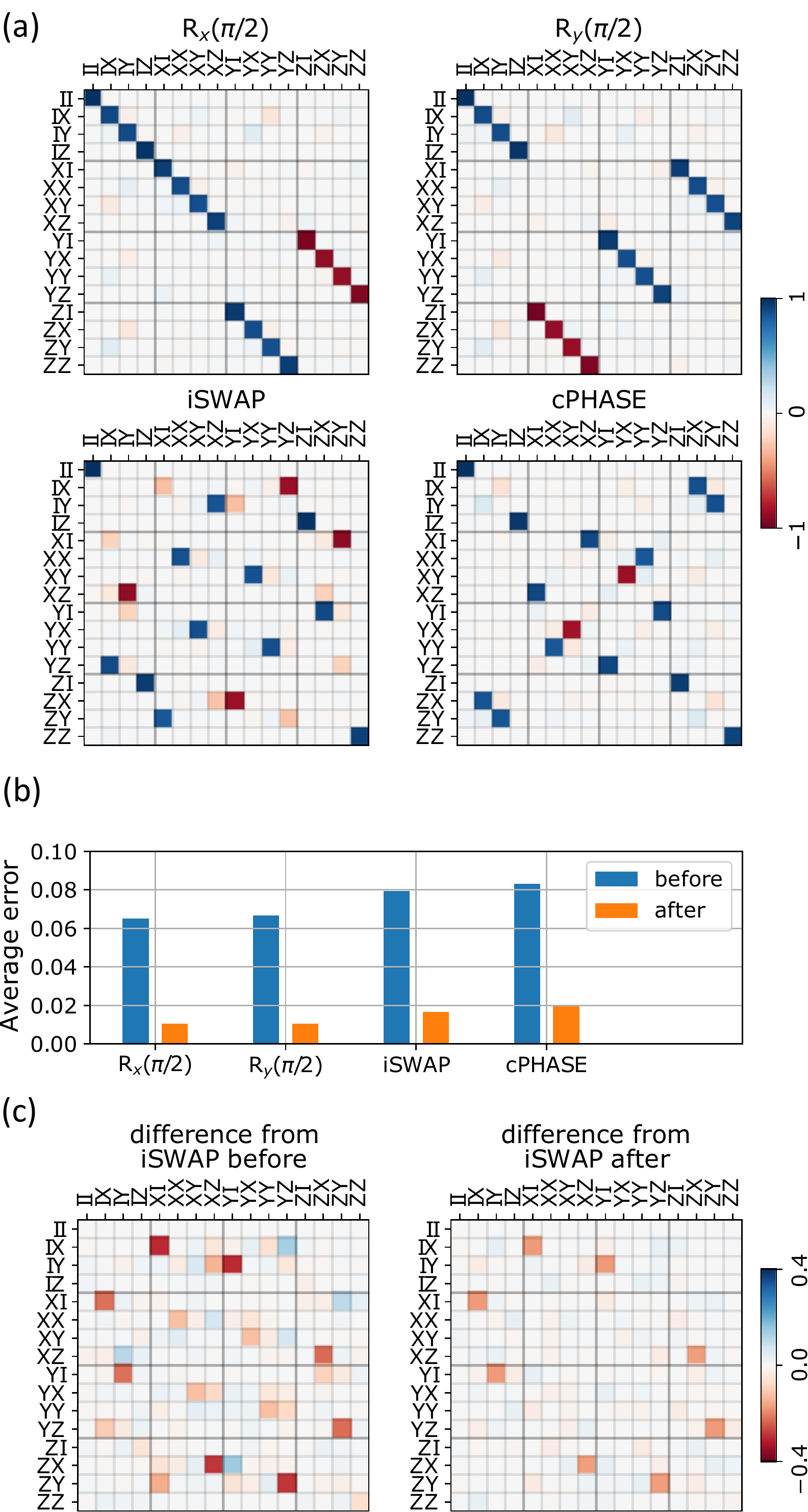}
\caption{(a) Pauli transfer matrices of the $\pi/2$ pulses on the control qubit, the iSWAP and the cPHASE gate, extracted in the first round of QPT. We do not show transfer matrices of ideal gates since they are essentially equal to the experimental matrices with all matrix elements rounded to $-1$, $0$ or $+1$. (b) Average errors of the extracted processes (calculated as $1-\overline{\mathcal{F}}$, where $\overline{\mathcal{F}}$ is the average fidelity of the process) before and after the iterative procedure. (c) Deviation of the extracted iSWAP Pauli transfer matrix from the ideal value, before and after the iterative procedure.}\label{fig:proctomo}
\end{figure}

We test this QPT procedure by applying it to the four basic pulses: $\mathrm{R}_x(\pi/2)$, $\mathrm{R}_y(\pi/2)$, iSWAP and cPHASE. The results are shown in Fig.~\ref{fig:proctomo}(a) where the extracted processes are represented by their \emph{Pauli transfer matrices} (i.e. matrices describing the action of the process on a density matrix expressed in the basis of Pauli operators \cite{Chow2012}).
While the obtained process matrices are relatively close to those of the ideal gates, there are also clear deviations. The fidelities of the extracted processes are around 0.935 for the single-qubit gates and 0.92 for the two-qubit gates. This unexpectedly large discrepancy is an artifact of the QPT method and highlights one of the drawbacks caused by the inaccessibility of the hidden qubit: In contrast to the standard setting with fully controllable qubits where the state preparation and tomography operations are single-qubit gates, here they are relatively complex sequences of both single-qubit and two-qubit gates. Consequently, state preparation and tomography errors due to imperfections in the gates are significantly more pronounced in a system with hidden qubits.

In principle, the effect of systematic gate errors can be compensated if the preparation and tomography operations are fully characterized. That is, if the processes $\mathcal{A}$ and $\mathcal{B}$ describing the non-ideal gate sequences in Eq.~(\ref{eq:QPTlstsq}) are known, the unknown process $\mathcal{X}$ can still be accurately extracted. However, we do not have any means of precisely characterizing the sequences without the very process tomography procedure we are trying to set up. To get around this circular dependency problem, we adopt an iterative method. Note that this particular technique for analyzing process tomography in a self-consistent manner may prove useful even for standard QPT with directly accessible qubits.

The QPT procedure applied to the gates $\mathrm{R}_x(\pi/2)$, $\mathrm{R}_y(\pi/2)$, iSWAP and cPHASE allows us to extract the process matrices $P = \{P_{x}, P_{y}, P_{\mathrm{iSWAP}}, P_{\mathrm{cPHASE}}\}$ of the four gates. In addition to the set of measurement data $D$, the obtained result $P$ also depends on the process matrices $P^{(\mathrm{SPAM})} = \{P^{(\mathrm{SPAM})}_{x}, P^{(\mathrm{SPAM})}_{y}, P^{(\mathrm{SPAM})}_{\mathrm{iSWAP}}, P^{(\mathrm{SPAM})}_{\mathrm{cPHASE}}\}$ assumed for the state preparation and measurement:
\[
  P = \mathrm{QPT}(D,P^{\mathrm{(SPAM)}})
\]
Alternatively, with the measured data fixed, the QPT procedure is a function mapping the set $P^{\mathrm{(SPAM)}}$ of four $16\times 16$ process matrices describing the gates used in the preparation and tomography sequences to the process matrices for the same four gates estimated by QPT.
Since the gates we are characterizing are the same as the ones forming the state preparation and tomography sequences, we would like to find a self-consistent set of process matrices for the gates, that is
\begin{equation}\label{eq:selfcons}
  P = \mathrm{QPT}(D,P).
\end{equation}
In other words, we wish to find a setting in which the process matrices describing the gates in the state preparation and tomography sequences are identical to the process matrices resulting from the QPT analysis.
Solving this equation for $P$ is generally difficult and is at the core of self-consistent tomography techniques proposed for robust characterization of gate sets \cite{Merkel2013,Greenbaum2015,Blume-Kohout2013}. These self-consistent approaches aim to correctly split the errors between the preparation and measurement gates and the process to be characterized.

If we wish to solve Eq.~(\ref{eq:selfcons}), it is important to notice that due to the absence of direct control over the hidden qubit, its solutions will be degenerate in the following sense: Both the initial state $|00\rangle\langle 00|$ and the native measurement operator $\sigma_z\otimes\mathbbm{1}$ commute with rotations of the hidden qubit around its $z$ axis. Hence, if we rotate all the gates in a sequence by an arbitrary angle $\varphi$ around this axis (i.e. we replace each gate $\mathcal{G}$ by $\mathcal{R}^{\dagger}\circ\mathcal{G}\circ\mathcal{R}$, where $\mathcal{R}$ is the superoperator $\mathcal{R}[\rho] = \mathrm{e}^{\mathrm{i}\varphi(\sigma_z\otimes\mathbbm{1})/2} \rho \mathrm{e}^{-\mathrm{i}\varphi(\sigma_z\otimes\mathbbm{1})/2}$), the measurement outcome remains unchanged. Thus, such rotations have no observable effect and so our process tomography by means of solving the self-consistency equation (\ref{eq:selfcons}) can determine the gate set only up to a rotation around the hidden qubit's $z$ axis.

However, the same reason that gives rise to this ambiguity -- the fact that a global rotation of all the used gates around the hidden qubit's $z$ axis has no effect on any experimental outcomes -- means we are free to assume an arbitrary value for this rotation parameter. Equivalently, since no drive pulses are applied directly to the hidden qubit, the phase of its rotating frame is a parameter which we can freely choose without having to make any changes to the applied pulse sequences.

We approximate a solution to Eq.~(\ref{eq:selfcons}) with an iterative algorithm. The ideal process matrices $P^{(\mathrm{ideal})} = \{P^{(\mathrm{ideal})}_x,P^{(\mathrm{ideal})}_y,P^{(\mathrm{ideal})}_{\mathrm{iSWAP}},P^{(\mathrm{ideal})}_{\mathrm{cPHASE}}\}$ for the four gates are a good starting approximation for $P$. When the sequence of iterates defined by
\begin{align*}
  P_0 &= P^{(\mathrm{ideal})}\\
  P_{i+1} &= (1-\lambda) P_{i} +
  \lambda \mathrm{QPT}(D,P_{i})
\end{align*}
converges, its limit is a solution to Eq.~(\ref{eq:selfcons}). The parameter $\lambda$ can be adjusted to improve the convergence properties. We use $\lambda = 0.1$ and find that while the iteration does not converge, the difference $P_i - f(P_i)$ (which may be seen as an indicator of how close $P_i$ is to a solution of Eq.~(\ref{eq:selfcons})) decreases for the first 30-40 iterations. Once it starts increasing, we stop the iteration and use the last value of $P_i$ as the result of our self-consistent QPT.

Note that apart from using the ideal gate process matrices as the starting point of the iteration, the procedure does not make use of the ideal gates in any way. There is thus no a priori reason to expect the process matrices resulting from the iterative method to be closer to the ideal gates than the ones obtained initially from standard QPT. Nevertheless, we find that the fidelities of the gates extracted by the iterative method are significantly improved (see Fig.~\ref{fig:proctomo}(b)), reaching approximately 0.99 for the single-qubit gates and 0.98 for the two-qubit gates, consistent with fidelities obtained by randomized benchmarking in Ref.~\cite{Ganzhorn2020}.

In Fig.~\ref{fig:proctomo}(c), we show the difference between the experimentally determined Pauli transfer matrix of the iSWAP gate and the ideal matrix. This difference is clearly reduced by the iterative procedure, indicating that the self-consistent approach works as intended -- the initial round of QPT assumes perfect state preparation and measurement and therefore all imperfections are lumped into the characterized gate. The self-consistent solution correctly takes into account that both the preparation and measurement gates as well as the characterized gate contribute to the overall error. Consequently, only a fraction of the total error is due to the characterized gate.

Note also that there are alternative methods to state or process tomography reconstruction \cite{Smolin2012,Lvovsky2004} which side-step the need to solve a general semidefinite programming problem. However, these rely on the measurement operators having the ideal form -- namely the set of 15 non-trivial two-qubit Pauli matrices. As we do not assume the measurement gates to be perfect, these methods are not directly applicable here.

\section{Conclusions}

We have discussed potential merits of systems in which some qubits are not directly controlled but are instead accessible only via other control qubits and two-qubit operations. Such devices can reach higher numbers of qubits for a fixed number of control and readout lines. As suggested by our analysis, this may pave a path to systems with higher quantum volumes in settings where control lines and the associated hardware are a limiting resource. On the topology discussed here (see Fig.~\ref{fig:networks}(f)), we find that for average decoherence-limited gate errors that are an order of magnitude lower than currently realized errors in superconducting qubits, $\Gamma \tau \lesssim 10^{-4}$, higher quantum volumes may be achieved by increasing the number of hidden qubits while keeping the number of control lines constant. While we focus on a specific topology with all hidden qubits connected directly to a neighboring control qubit, other topologies may lead to a bigger advantage for a fixed number of control lines. The advantage of hidden qubit architectures may be even more pronounced if hidden qubits can reach longer coherence times due to the lack of coupling to direct control and readout lines.

To demonstrate the operation of such a system and highlight some of the challenges and ways to address them, we have experimentally characterized a device with one control and one hidden qubit. We have shown that despite the absence of a direct drive on the qubit as well as the lack of a read-out, we can calibrate a gate set which gives us full control over the system, allowing us to perform quantum process tomography. This becomes possible by implementing both iSWAP and cPHASE gates based on a parametrically driven tunable coupler side-by-side. These gates form, together with full control of a single qubit, a complete set of gates.

To address the problem of state preparation and measurement errors in the tomography procedure, we have implemented an iterative algorithm that extracts the process matrices for the calibrated gate set in a self-consistent manner.

\section{Acknowledgments}

We thank Stephan Paredes, Andreas Fuhrer, Matthias Mergenthaler, Peter M\"{u}ller and Clemens M\"{u}ller for insightful discussions and the quantum team at IBM T. J. Watson Research Center, Yorktown Heights for the provision of qubit devices. We thank R.
Heller and H. Steinauer for technical support.
Fabrication of samples was financially supported by the ARO under contract W911NF-14-1-0124,
M. W. acknowledges funding by the European Commission Marie Curie ETN project
QuSCo (Grant Nr. 765267), and M. P. and G. S. by the European FET-OPEN
project Quromorphic (Grant Nr. 828826)

\bibliographystyle{apsrev4-1}
\bibliography{main.bbl}

\appendix

\section{Quantum volume calculation}\label{app:VQ}
To compare the fully controlled square grid configuration from Fig.~\ref{fig:networks}(d) with the network containing hidden qubits as in Fig.~\ref{fig:networks}(f), we use quantum volume as defined in Ref.~\cite{Cross2019}. To calculate it, we estimate the typical circuit depth $d(N)$ which is in a certain sense "achievable" in a network with $N$ qubits. The circuit depth is defined as the number of circuit layers, where each layer consists of single-qubit gates followed by two-qubit gates among $N/2$ disjoint qubit pairs, while the exact meaning of "achievable" differs between references. We adopt the simple definition from \cite{Moll2018} where $d(N)$ is the number of layers at which the expected overall error probability reaches some fixed threshold $\varepsilon$. At the same time, we assume error probabilities combine additively and if the typical total error per circuit layer is $\varepsilon_{\mathrm{l}}(N)$, we approximate the achievable circuit depth as $d(N) = \varepsilon/\varepsilon_{\mathrm{l}}(N)$.
In a system without all-to-all connectivity, each two-qubit gate typically needs to be decomposed into several gates between nearest neighbors. Each circuit layer will therefore contain many more two-qubit gates than single-qubit gates. Since the error probability of single-qubit gates is typically lower than that of two-qubit gates, the error probability $\varepsilon_{\mathrm{l}}(N)$ will be dominated by the two-qubit gates. We therefore assume perfect single-qubit gates.

The decomposition of a set of $N/2$ two-qubit gates into nearest neighbor operations will consist of $n_{\mathrm{g}}\ge N/2$ gates. These can be split into groups acting on disjoint pairs of qubits and each group can be applied simultaneously in a single time step of length $\tau$. The overall procedure will therefore take $n_{\mathrm{s}}\le n_{\mathrm{g}}$ time steps.

In our simplified analysis, we assume the errors are dominated by decoherence mechanisms and the total error per circuit layer can be approximated as $N n_{\mathrm{s}}(N)\Gamma\tau$, where $\Gamma$ is an effective error rate per qubit.

The circuit depth is
\[
  d(N) = \frac{\varepsilon}{
  N n_{\mathrm{s}}(N) \Gamma \tau
  }.
\]
and the quantum volume is
\[
  \log_2 V_Q(N) = \mathrm{min}\left(\frac{\varepsilon}{
  N n_{\mathrm{s}}(N) \Gamma \tau
  },N\right).
\]

Since hidden qubits may be better isolated from certain dissipation and dephasing channels and could therefore reach longer coherence times than control qubits, we also consider an alternative setting where the two types of qubits have different associated probabilities of decoherence-induced errors. We assume the hidden qubits experience a lower error rate $\Gamma$ per qubit while control qubits may still be subject to a higher error rate $\Gamma^{(\mathrm{c})}$. This can be taken into account simply by replacing the term $N n_{\mathrm{s}}(N) \Gamma\tau$ in the denominator in the equation for $V_Q(N)$ with
\[
  n_{\mathrm{s}}(N) (N_{\mathrm{c}}
  \Gamma^{(\mathrm{c})} +
  N_{\mathrm{h}} \Gamma)
  \tau,
\]
where $N_{\mathrm{c}}$ and $N_{\mathrm{h}}$ are the numbers of control and hidden qubits, respectively.

While we have used a simplistic error model, the analysis we are about to describe also gives us access to quantities such as the typical number of gates of each type (swap gates or entangling gates) used in the decomposition of a random circuit layer. It can therefore be used in a straightforward manner even for models where the total error probability depends on the number of gates (as it would for control errors) and is not equal for all gates.

To calculate $V_Q$, we adopt a specific algorithm for decomposing an arbitrary set of gates between disjoint pairs of qubits into nearest neighbor gates. We then sample random choices of the qubit pairs to obtain the average numbers of gates $n_{\mathrm{g}}(N)$ and time steps $n_{\mathrm{s}}(N)$. This algorithm -- an instance of a transpiler -- is described in more detail below. We have not compared its performance with other, more sophisticated transpiler algorithms \cite{Cross2019}. While our quantum volume estimates may be lower than what could be achieved with highly optimized transpilers, we believe that a meaningful comparison can still be made between grids with hidden qubits and fully controlled ones.

\begin{figure}
\includegraphics[width=8.0cm]{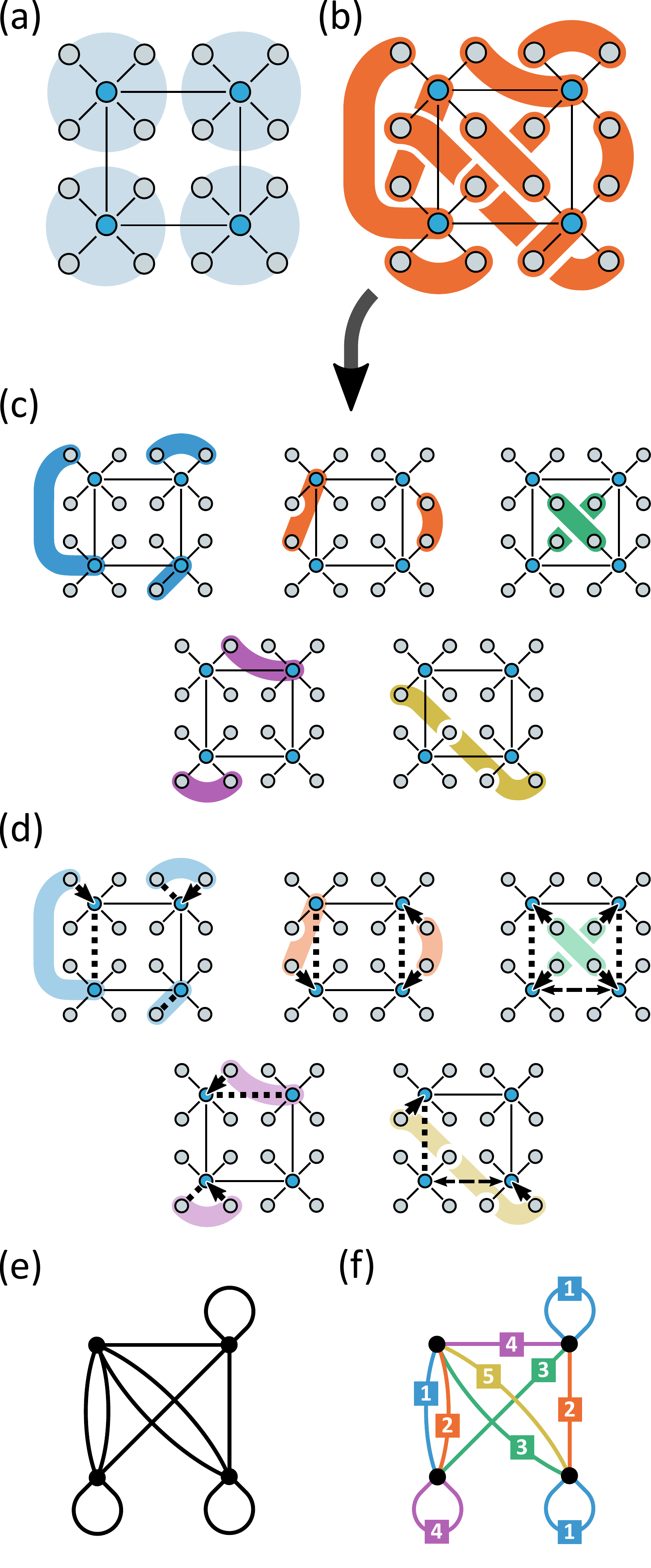}
\caption{(a) Example of a network with $k=2$ and $h=4$ hidden qubits per grid site. The circles indicate the \emph{grid groups} of qubits. (b) One instance of a grouping of the 20 qubits into 10 pairs which are to be coupled by two-qubit gates. (c) Decomposition of the pairs into groups $G_1,\ldots,G_5$ such that for each $G_i$ no two qubit pairs belonging to it share a grid group. (d) Implementation of the gates between the qubit pairs in each group as a combination of swaps between hidden and control qubits (solid arrows), swaps between control qubits (dashed arrows) and entangling gates (dotted lines). (e) Multigraph corresponding to the qubit pairing from (b) and its edge coloring (f) which results in the decomposition from (c).}
\label{fig:graphs}
\end{figure}

Let us assume a grid size of $k\times k$ control qubits, each connected to $h$ hidden qubits as shown in Fig.~\ref{fig:networks}(f) (the fully controllable system has $h=0$). We will call every controlled qubit with its $h$ associated hidden qubits a \emph{grid group}. In total, we have $N = (h+1)k^2$ qubits in $k^2$ grid groups. A simple example with $k=2$ and $h=4$ is shown in Fig.~\ref{fig:graphs}(a). The gate decomposition algorithm is based on moving the paired qubits (Fig.~\ref{fig:graphs}(b)) around the grid by SWAP operations until they are next to each other, then applying the desired two-qubit gate and finally swapping them back to their original locations. For simplicity, we assume that the targeted two-qubit operation counts as a single gate and we will not distinguish between its error probability and those of the SWAP gates. A more accurate analysis would take into account the decomposition of the entangling operation into a library of 'hardware-native' gates \cite{Bullock2003,Shende2004}.

The overall procedure to implement the gates between the desired pairs of qubits is accomplished in the following steps:
\begin{enumerate}
  \item{Split the $(h+1)k^2/2$ qubit pairs into groups $G_1,G_2,\ldots$ such that for each $G_i$ no two pairs belonging to it share a grid group (see Fig.~\ref{fig:graphs}(c)).}
  \item{Take one of these groups and swap all its hidden qubits with their corresponding controlled qubits, except for pairs where both qubits belong to the same grid group. In this case, swap only one of them if both are hidden, otherwise do nothing. This step is depicted by the solid arrows in Fig.~\ref{fig:graphs}(d)}
  \item{Permute the controlled qubits using SWAP gates (as depicted by the dashed arrows in Fig.~\ref{fig:graphs}(d)) in such a way that the paired qubits end up next to each other.}
  \item{Apply two-qubit gates between the paired qubits (shown by dotted lines in Fig.~\ref{fig:graphs}(d)).}
  \item{Undo swaps from step 3.}
  \item{Undo swaps from step 2. If there are more groups $G_i$ to be processed, go to step 2. Otherwise the process is finished.}
\end{enumerate}

To realize step 1, we construct a graph with vertices representing the grid groups and with an edge for each pair of qubits belonging to the two grid groups (since multiple pairs of qubits can be shared between two grid groups, this means there may be more than one edge between two vertices of the graph, making it in graph theory language a \emph{multigraph}). Such a multigraph for the example pairing from Fig.~\ref{fig:graphs}(b) is shown in Fig.~\ref{fig:graphs}(e). Splitting the qubit pairs into the groups $G_1,G_2,\ldots$ is equivalent to splitting the edges into sets where edges in the same set do not share vertices, in other words to \emph{edge coloring} of the multigraph. One possible coloring for our example is illustrated in Fig.~\ref{fig:graphs}(f). This results in the groups from Fig.~\ref{fig:graphs}(c). We find these colorings using the \texttt{networkx} Python module. As an aside, since each vertex of the multigraph clearly has degree $h+1$, a classic theorem on multigraph edge coloring by Shannon \cite{Shannon1949b} allows us to upper-bound the required total number of pair groups $G_i$ by $\frac{3}{2}(h+1)$.

In step 2, we take one of the groups $G_i$ and move its qubits onto the grid vertices, such that we can start moving them towards each other in the next step. The condition we placed on the groups $G_i$ (that qubit pairs within $G_i$ do not share a grid group) ensures that there are no collisions in this step.

Step 3 essentially means picking a certain permutation of the grid qubits that brings the pairs to be coupled next to each other and then realizing it by means of nearest neighbor swaps. The permutation is obviously not unique since there is a lot of freedom in choosing at which nearest neighbor sites the individual qubit pairs will meet. We choose the option which minimizes the sum of the $L^{1}$ distances by which the individual qubits need to be moved from their original positions to their destination sites. Note that this summed distance is \emph{not} equal or even directly related to the length of the paths which will actually be taken by the qubits nor to the total number of SWAP gates used. This is because the algorithm chosen to realize the permutation (described below) does not send each qubit via the shortest path. We merely choose the sum of the shortest path distances as a convenient heuristic to roughly judge the suitability of each qubit rearrangement. This type of constrained assignment problem, which asks to map each qubit pair to some nearest neighbor pair of sites such that the sites chosen for different qubit pairs do not overlap and that the sum of distances is minimized, can be formulated as an integer linear programming problem. We solve this problem numerically using the \texttt{cvxopt} Python package.

To realize the chosen permutation using nearest neighbor swaps, we use a method described in \cite{Alon1994} where an arbitrary permutation is decomposed into three permutations, the first and last being column-wise and the middle one row-wise, i.e. qubits are permuted only within individual columns or rows. A permutation within a column or a row is seen as a simple sorting task whose implementation as a series of nearest neighbor swaps is given for instance by the bubble sort algorithm. Importantly, the swaps within distinct columns or rows can be realized in parallel.

\section{Tune-up of the gate set}\label{app:gate_set_tuneup}

Here we summarize the procedure used to calibrate our gate set. Parts of the process are similar to that described in \cite{Ganzhorn2020}. Calibration of single-qubit gates on the control qubit is standard and we do not describe it here. The main differences from \cite{Ganzhorn2020} are the methods used to measure the two-qubit phase of the cPHASE gate (here we make use of a spin echo measurement) and the necessity to treat the iSWAP phase tune-up differently because the hidden qubit is not directly accessible.

As in the main text, $\mathrm{R}_j(\varphi)$ will denote single-qubit rotations around axis $j$ by an angle $\varphi$. In addition to $j\in\{x,y\}$, we also denote rotations around an arbitrary axis in the $xy$ plane with azimuthal angle $\theta$ by $j = \theta$. The gates whose parameters we are adjusting in the tune-up procedure will be called SW and CP (to distinguish them from the ideal or already tuned-up gates iSWAP and cPHASE).

All sequences end with a qubit state measurement which we do not explicitly write out.

\subsection{Calibrating iSWAP pulse length and frequency}
With sequences
\begin{equation}
  \mathrm{SW}^n. \mathrm{R}_x(\pi), \tag{S1}
\end{equation}
where the superscript $n$ denotes repetition of the gate $n$ times, we prepare the control qubit in its excited state, then apply the candidate SW gate and measure the control qubit's final excitation $p_e$. In addition to a single SW gate ($n = 1$), we also use sequences with multiples of them ($n = 3$ and $n = 5$). This repetition increases the sensitivity of the measurement to deviations of SW from the ideal iSWAP gate.

\begin{figure}
\includegraphics[width=8.5cm]{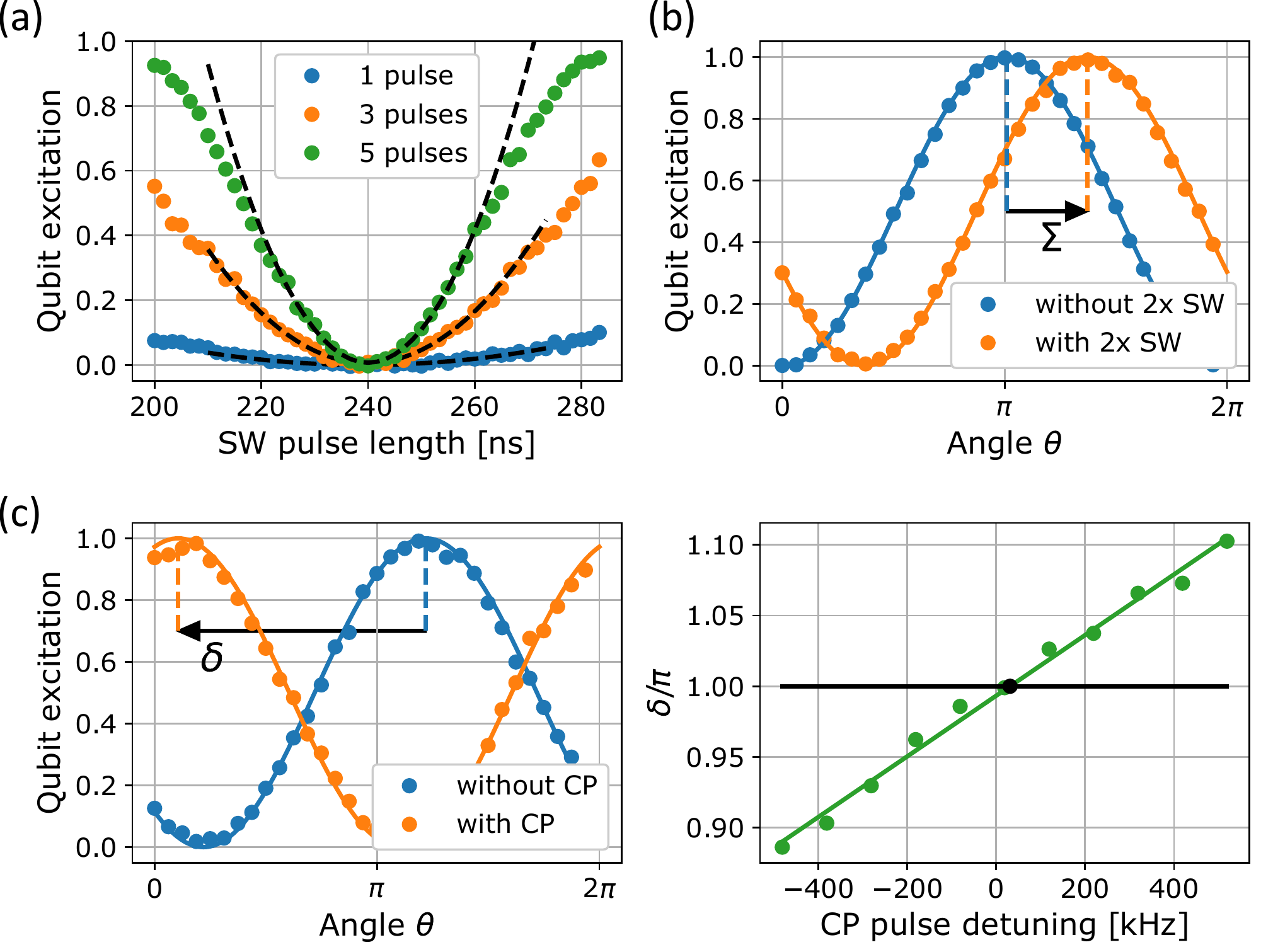}
\caption{Experimental data from typical gate calibration measurements. (a) Calibration of iSWAP pulse length. The final excitation probability of the control qubit reaches its minimum at the optimal pulse length. We extract this optimum by fitting a quadratic function in the vicinity of the minimal value. The pulse lengths obtained from measurements with 1, 3 or 5 repetitions of the pulse are consistent. (b) Calibration of iSWAP single-qubit phases. The final control qubit excitation after the Ramsey sequence shows a cosine oscillation as a function of the phase $\theta$ of the second $\pi/2$ pulse. The shift of this curve between Ramsey sequences with and without two SW pulses give us the phase sum $\Sigma=\gamma_1+\gamma_2$. (c) Calibration of cPHASE frequency. The two-qubit phase $\delta = \gamma_{11}-\gamma_{01}-\gamma_{10}$ is measured with a Ramsey sequence utilizing spin echo. As in (b), the phase $\theta$ of the final $\pi/2$ pulse is varied and the shift of the obtained cosine dependence between sequences with and without the CP pulse is the desired phase $\delta$. We perform this measurement for varying detuning of the CP pulse and find a dependence which is well approximated by a linear function. Using this fit, we obtain the optimal detuning where $\delta=\pi$.}
\label{fig:gatecal}
\end{figure}

After running this experiment for a range of lengths of the SW pulse, we find the length which leads to minimal $p_e$ (see Fig.~\ref{fig:gatecal}(a)). We then perform the same experiment but this time for different detunings of the SW pulse. Again, we choose the detuning which minimizes $p_e$. If needed, we repeat the cycle consisting of length and detuning optimization several times until the parameters have converged.

\subsection{Calibrating iSWAP single-qubit phases}

Assuming the population transfer of the SW gate has already been optimized and is perfect, the unitary describing the gate has the form
\[
  \left(\begin{array}{cccc}
  1 & 0 & 0 & 0\\
  0 & 0 & \mathrm{e}^{\mathrm{i}\gamma_1} & 0\\
  0 & \mathrm{e}^{\mathrm{i}\gamma_2} & 0 & 0\\
  0 & 0 & 0 & \mathrm{e}^{\mathrm{i}\gamma_3}
  \end{array}\right),
\]
in the basis $\{|00\rangle, |10\rangle, |01\rangle, |11\rangle\}$. Here the first of the two qubits is the control and the second the hidden one.

In the standard setting where both qubits are directly controllable, we would implement iSWAP by following this gate with rotations of the two qubits around their individual $Z$ axes (which is usually done virtually, i.e. by shifting the qubits' rotating reference frames \cite{McKay2017}). If the virtual $Z$ phases are $\delta_1$ and $\delta_2$, the unitary above is transformed into
\begin{equation}\label{eq:swapWithFrameChange}
  \left(\begin{array}{cccc}
  1 & 0 & 0 & 0\\
  0 & 0 & \mathrm{e}^{\mathrm{i}(\gamma_1-\delta_1)} & 0\\
  0 & \mathrm{e}^{\mathrm{i}(\gamma_2-\delta_2)} & 0 & 0\\
  0 & 0 & 0 & \mathrm{e}^{\mathrm{i}(\gamma_3-\delta_1-\delta_2)}
  \end{array}\right).
\end{equation}
With only two degrees of freedom to tweak, we cannot in general adjust all of the matrix elements to achieve the ideal iSWAP values (i.e. $\mathrm{e}^{\mathrm{i}(\gamma_1-\delta_1)}=\mathrm{e}^{\mathrm{i}(\gamma_2-\delta_2)}=\mathrm{i}$ and $\mathrm{e}^{\mathrm{i}(\gamma_3-\delta_1-\delta_2)}=1$) unless the phases $\gamma_{1,2,3}$ are related such that $\mathrm{e}^{\mathrm{i}(\gamma_1+\gamma_2)} = -\mathrm{e}^{\mathrm{i}\gamma_3}$. In practice, because of the coupling of $\ket{11}$ to neighbouring higher-excited states this relation does not hold exactly but is a good approximation. We have $\gamma_3 + \pi =  \gamma_1 + \gamma_2 + \beta$, where $\beta$ is small. We typically set the frame shifts $\delta_1$ and $\delta_2$ such that $\gamma_3 - \delta_1 - \delta_2 = 0$ and $\gamma_{1,2} - \delta_{1,2} = \pi/2 - \beta/2$, a setting that leads to a lower error than the alternative $\gamma_{1,2}-\delta_{1,2} = \pi$ and $\gamma_3 - \delta_1 - \delta_2 = \beta$ (see Ref.~\cite{Ganzhorn2020}). This leads to the unitary
\begin{equation}\label{eq:iSWAPwithZZ}
  \left(\begin{array}{cccc}
  1 & 0 & 0 & 0\\
  0 & 0 & \mathrm{i}\,\mathrm{e}^{-\mathrm{i}\beta/2} & 0\\
  0 & \mathrm{i}\,\mathrm{e}^{-\mathrm{i}\beta/2} & 0 & 0\\
  0 & 0 & 0 & 1
  \end{array}\right),
\end{equation}
where $\mathrm{e}^{\mathrm{i}\beta}$ is close to $1$.

In the setting where one of the qubits is hidden, there area a few crucial differences. Since excitations can be created and measured only in the control qubit, any Ramsey-type measurement of the phases $\gamma_{1,2,3}$ needs to involve an even number of swaps (the excitation swapped into the hidden qubit is swapped back when measured). Therefore the phases $\gamma_1$ and $\gamma_2$ are not observable individually but only via their sum $\Sigma\equiv\gamma_1+\gamma_2$.

Another way to see this is the following: Neglecting decoherence, the result of any experiment involving the setup with one control and one hidden qubit is described by the probabilities $\mathrm{Tr}(P_0 U_n\ldots U_1\rho_0 U_1^{\dagger}\ldots U_n^{\dagger})$ and $\mathrm{Tr}(P_1 U_n\ldots U_1\rho_0 U_1^{\dagger}\ldots U_n^{\dagger})$, where $P_j$ are the projections onto the computational states of the control qubit $P_j = |j\rangle\langle j|\otimes\mathbbm{1}$, $\rho_0$ is the initial state $|00\rangle\langle 00|$ and $U_k$ the unitaries describing the individual gates.

If $R_z$ is a rotation of the control qubit around its $Z$ axis by an arbitrary angle $\delta$, the probabilities above do not change when we perform a replacement $X\to R_z^{\vphantom{\dagger}} X R_z^{\dagger}$ on all the operators. Moreover, all the unitaries except the swap operations commute with $R$ and therefore the outcome of the experiment remains unchanged when we replace just the swaps among the $U_j$ operators by $R_z^{\vphantom{\dagger}} U_j R_z^{\dagger}$. This replacement is equivalent to changing $\gamma_1\to\gamma_1+\delta$ and $\gamma_2\to\gamma_2-\delta$. Hence, the outcome probabilities $p_{0,1}$ as a function of the parameters $\gamma_{1,2,3}$ satisfy
\[
  p_{0,1}(\gamma_1,\gamma_2,\gamma_3) =
  p_{0,1}(\gamma_1+\delta,\gamma_2-\delta,\gamma_3)
\]
for arbitrary $\delta$. Choosing specifically $\delta = \gamma_2$, we see that $p_{0,1}(\gamma_1,\gamma_2,\gamma_3) = p_{0,1}(\gamma_1+\gamma_2,0,\gamma_3)$. This result shows that the (only) observable quantities in this system depend on $\gamma_1$ and $\gamma_2$ only via $\Sigma = \gamma_1+\gamma_2$.

While this implies that we can indeed measure only the sum and not the individual parameters $\gamma_{1,2}$, it also means we do not need to. Any experiment using iSWAP gates implemented as the unitary given in Eq.~(\ref{eq:swapWithFrameChange}) will be equivalent to one where the iSWAP has the proper form from Eq.~(\ref{eq:iSWAPwithZZ}) as long as
\begin{equation}\label{eq:phaserel}
\gamma_1+\gamma_2-\delta_1-\delta_2 = \pi - \beta.
\end{equation}
We are free to choose the rotating frame shifts $\delta_1$ and $\delta_2$ as long as they together satisfy this equation.

This may sound suspicious -- we might think that since no pulses are applied to the hidden qubit, the frame change $\delta_2$ does not affect the experiment in any way (in which case $\delta_1$ would have to be irrelevant as well). However, we need to bear in mind that the parametric drive inducing the iSWAP process is defined in a rotating frame which is derived from the individual qubits' frames (loosely speaking as their difference). Therefore $\delta_2$ enters into the experimental parameters via the shift of the parametric drive frame which has to be $\delta_1 - \delta_2 \equiv \delta_p$.

In our calibration measurement, we determine $\Sigma = \gamma_1+\gamma_2$ in a Ramsey-type measurement consisting of the pulse sequences
\begin{align}
  \mathrm{R}_\theta(\pi/2) . \mathrm{SW} .
  \mathrm{SW} . & \mathrm{R}_x(\pi/2) \tag{S2}
  \label{pseq:iswap1qphases11}\\
  \mathrm{R}_\theta(\pi/2) . & \mathrm{R}_x(\pi/2). \tag{S3}
  \label{pseq:iswap1qphases12}
\end{align}
By measuring the final qubit excitation as a function of the angle $\theta$, we get an oscillatory dependence whose phase shift gives us the orientation of the final qubit Bloch vector in the $xy$ plane. The difference in the Bloch vector orientation between the sequences with and without the SW pulses directly gives us the phase $\Sigma = \gamma_1+\gamma_2$ (see Fig.~\ref{fig:gatecal}(b)).

Repeating (\ref{pseq:iswap1qphases11}) and (\ref{pseq:iswap1qphases12}) with the hidden qubit prepared in its excited state, i.e. using the sequences
\begin{align}
  \mathrm{R}_\theta(\pi/2) . \mathrm{SW} .
  \mathrm{SW} . \mathrm{R}_x(\pi/2) . \mathrm{SW} .
  & \mathrm{R}_x(\pi)\tag{S4}
  \label{pseq:iswap1qphases21}\\
  \mathrm{R}_\theta(\pi/2) . \mathrm{R}_x(\pi/2) .
  \mathrm{SW} . & \mathrm{R}_x(\pi), \tag{S5}
  \label{pseq:iswap1qphases22}
\end{align}
yields the phase $\gamma_3-\gamma_1-\gamma_2 = \beta-\pi$. We can then correctly set the frame changes, for instance by arbitrarily choosing $\delta_2 = 0$ and then calculating $\delta_1=\delta_p$ from Eq.~(\ref{eq:phaserel}).

\subsection{Calibrating cPHASE pulse length}

Similarly to the iSWAP length calibration, we use the sequences
\begin{equation}
  \mathrm{CP}^n . \mathrm{R}_x(\pi) . \mathrm{iSWAP} .
  \mathrm{R}_x(\pi). \tag{S6}
\end{equation}
The gates $\mathrm{R}_x(\pi) . \mathrm{iSWAP} .
\mathrm{R}_x(\pi)$ prepare the state $|11\rangle$, after which the candidate CP gate is applied ($n = 1$, $3$ or $5$ times). This measurement is repeated for a range of lengths of the CP pulse to find the one which maximizes final control qubit excitation $p_e$.

\subsection{Calibrating cPHASE pulse frequency}

Detuning the CP pulse changes the phase accumulated by the $|11\rangle$ state. To make it equivalent (up to single-qubit rotations) to a cPHASE gate, the phases $\gamma_{01}$, $\gamma_{10}$ and $\gamma_{11}$ accumulated by the computational states (relative to the $|00\rangle$ state) must satisfy
\[
  \gamma_{11} - \gamma_{01} - \gamma_{10} = \pi.
\]
We measure $\gamma_{11} - \gamma_{01} - \gamma_{10}$ this combination of phases using the two sequences
\begin{align*}
  & \mathrm{R}_\theta(\pi/2) . \mathrm{CP} .
  \mathrm{FLIP} . \mathrm{CP} . \mathrm{R}_x(\pi/2)\tag{S7}\\
  & \mathrm{R}_\theta(\pi/2) . \mathrm{FLIP} .
  \mathrm{R}_x(\pi/2)\tag{S8}\\[3mm]
  & \text{where }
  \mathrm{FLIP} :=
  \mathrm{iSWAP} . \mathrm{R}_x(\pi) . \mathrm{iSWAP} .
  \mathrm{R}_x(\pi).
\end{align*}
These Ramsey-type measurements, where FLIP flips the states of both qubits, can be interpreted as a spin echo experiment measuring the difference between the phases induced by the CP pulse on the control qubit when the hidden qubit is in the ground or the excited state. This difference is exactly $\delta\equiv\gamma_{11} - \gamma_{01} - \gamma_{10}$. The second sequence without the CP pulse serves as a reference to subtract phases induced by the iSWAP gate (see Fig.~\ref{fig:gatecal}(c)).

We perform this experiment for a range of detunings of the CP pulse and choose the value for which the measured phase $\delta\equiv\gamma_{11}-\gamma_{01}-\gamma_{10}$ is closest to $\pi$.

\subsection{Calibrating single-qubit phases induced by cPHASE}

To adjust the single-qubit phases $\gamma_{01}$ and $\gamma_{10}$ to zero, we first measure them and then compensate them by shifting the qubits' reference frames.

The phase induced on the control qubit can be measured in a straightforward Ramsey experiment:
\begin{align}
  \mathrm{R}_\theta(\pi/2) . \mathrm{CP} . &
  \mathrm{R}_x(\pi/2)\tag{S9}\\
  \mathrm{R}_\theta(\pi/2) . & \mathrm{R}_x(\pi/2)\tag{S10}
\end{align}
For the hidden qubit, we need to add iSWAP gates after the first and before the second $\pi/2$ pulse. In this case, the reference measurement without the CP pulse is used to subtract any potential phases induced by the iSWAP gate:
\begin{align}
  \mathrm{R}_\theta(\pi/2) . \mathrm{SW} .
  \mathrm{CP} . \mathrm{SW} . & \mathrm{R}_x(\pi/2)\tag{S11}\\
  \mathrm{R}_\theta(\pi/2) . \mathrm{SW} .
  \mathrm{SW} . & \mathrm{R}_x(\pi/2)\tag{S12}
\end{align}

\end{document}